\documentclass{osa-article}
\pdfoutput=1
\usepackage{amsmath}
\usepackage{amssymb}
\usepackage[numbers,sort&compress]{natbib}
\usepackage{graphicx}
\usepackage{epstopdf}
\usepackage{subfigure}
\journal{osajournal}

\articletype{Research Article}

\begin{document}

\title{Diffractive deep neural network based adaptive optics scheme for vortex beam in oceanic turbulence}
\author{Haichao Zhan,\authormark{1} Le Wang,\authormark{1} Wennai Wang,\authormark{2} and Shengmei Zhao,\authormark{1,2,*}}

\address{\authormark{1}Institute of Signal Processing and Transmission, Nanjing University of Posts and Telecommunications(NUPT), Nanjing 210003, China\\
\authormark{2}Key Lab of Broadband Wireless Communication and Sensor Network Technology (Nanjing University of Posts and Telecommunications), Ministry of Education, Nanjing 210003, China}
\email{\authormark{*}zhaosm@njupt.edu.cn} 

\begin{abstract}
Vortex beam carrying orbital angular momentum (OAM) is disturbed by oceanic turbulence (OT) when propagating in underwater wireless optical communication (UWOC) system. Adaptive optics (AO) is used to compensate for distortion and improve the performance of the UWOC system. In this work, we propose a diffractive deep neural network (DDNN) based AO scheme to compensate for the distortion caused by OT, where the DDNN is trained to obtain the mapping between the distortion intensity distribution of the vortex beam and its corresponding phase screen representating OT. The intensity pattern of the distorted vortex beam obtained in the experiment is input to the DDNN model, and the predicted phase screen can be used to compensate the distortion in real time. The experiment results show that the proposed scheme can extract quickly the characteristics of the intensity pattern of the distorted vortex beam, and output accurately the predicted phase screen. The mode purity of the compensated vortex beam is significantly improved, even with a strong OT. Our scheme may provide a new avenue for AO techniques, and is expected to promote the communication quality of UWOC system.
\end{abstract}

\section{Introduction}

In recent years, underwater wireless optical communication (UWOC) has attracted a lot of attentions due to its high data rate, high security, high bandwidth, low latency, and low cost \cite{oubei20154,song2017experimental,7982610,zhao2020long}. UWOC has many applications in the rapidly growing human underwater activities, such as submarines, autonomous underwater vehicles and unmanned underwater vehicles \cite{9393797,Xu:19}. In order to meet the increasing demand for data transmission in military, civilian and commercial, orbital angular momentum (OAM) is introduced into UWOC \cite{ZHAN2021166990}. In 1992, Allen \emph{et.al.} proved that OAM mode can be characterized by a spatial wave-function with a helical phase $\text{exp}\left( i\ell \theta  \right)$ \cite{PhysRevA.45.8185}, where $\ell$ is an arbitrary integer named topological charge, representing the number of $2\pi $ phase shifts across the beam, and $\theta $ corresponds to the azimuthal angle. The different OAM modes carried by the beam are orthogonal, which provides a new degree of freedom for the multiplexing of OAM beams, and increases transmission capacity and spectral efficiency \cite{ISI:000305905000019}. Unfortunately, the spiral wavefront structure of OAM mode is easily distorted by oceanic turbulence (OT), which aggravates crosstalk between OAM modes and affects the communication quality of the UWOC system \cite{8966342}. Therefore, adaptive optics (AO), as a technique that can reduce the distortion caused by turbulence, is applied to the UWOC system \cite{Li:14}.

Generally, Shack-Hartmann (SH) wavefront correction method needs an SH wavefront sensor (WFS) to obtain distortion information, and a deformable mirror to correct wavefront distortion. But, the existence of the phase singularity causes the SH wavefront correction method to be unable to reconstruct the spiral wavefront of the OAM mode well \cite{Hu:19}. Accordingly, the AO method without WFS has received a lot of attention, such as, Gerchberg-Saxton (GS) algorithm \cite{Chang:17}, stochastic parallel gradient descent (SPGD) algorithm \cite{Xie:15}, hybrid input-output algorithm \cite{Yin:18}, genetic algorithm \cite{Tehrani:15}. However, these methods need to be iterated, which consumes more time and is easy to fall into the trouble of local optimization. Recently, the AO method without WFS combined with deep learning \cite{LeCun2015} has become a research hotspot. Zhai \emph{et.al.} proposed an AO method based on convolutional neural network (CNN) \cite{Zhai:20}. The CNN trained the corresponding relationship between the distortion intensity recorded by a charge-coupled device (CCD) camera and the first 20 Zernike coefficients representing the turbulence phase, then output the predicted Zernike coefficients to compensate for distortion. Later, Lu \emph{et.al.} designed a jointly-trained CNN-based OAM recognition method, which implicitly uses an upsample model as a substitute for a one-shot AO system \cite{Lu:20}. Upsampling and recognition share the same CNN backbone, and then use transposed convolution to restore the distortion intensity image.

In 2018, Lin \emph{et.al.} introduced an all-optical deep learning network framework in which the neural work is composed of multiple layers of diffractive surfaces, which is called a diffractive deep neural network (DDNN) \cite{Lin:18}. DDNN can realize various functions based on deep learning through the all-optical passive diffraction layer, so it has attracted a lot of interests. Zhao \emph{et.al.} proposed an OAM mode detection method based on DDNN, and tested the recognition performance of three types of networks: amplitude-only, phase-only, and hybrid type \cite{Zhao:19}. Wang \emph{et.al.} utilized DDNN to realize the logic operation of OAM mode, which improved the parallel processing capability and enhanced the logic robustness due to the infinity and orthogonality of OAM modes \cite{Wang:21}. Huang \emph{et.al.} used DDNN's light field control ability to realize the signal processing of the vortex beam, including OAM-shift-keying (OAM-SK), OAM multiplexing and demultiplexing, and OAM-mode switching \cite{PhysRevApplied.15.014037}. These researches provide references for the realization of AO based on DDNN.

In this paper, we propose a DDNN-based AO compensation scheme, where DDNN is trained to learn the relationship between the intensity pattern of the vortex beam and the phase screen simulating OT by numerical simulations. We exploit CCD to record the intensity pattern of the vortex beam that is distorted by OT in experiment, and input the distorted intensity image to the trained DDNN to obtain a predicted phase screen, then use the predicted phase screen to compensate the distorted vortex beam. We demonstrate the reliability of the proposed AO scheme, and compare the compensation performance with the previous CNN-based AO schemes and the conventional GS algorithm under different OT intensities.

\section{Theory}

\subsection{DDNN-based AO scheme}

The concept of the AO compensation scheme based on DDNN is sketched in Fig. 1. In Fig. 1(a), a vortex beam passes through a phase screen that simulating OT, its wavefront is distorted, and its perfect doughnut-like intensity distribution is also destroyed. The distorted intensity distribution image and phase screen are input to the trained DDNN to obtain the phase screen representating the distortion. Then, the phase screen output by DDNN is reversed phase operation to form a compensation phase screen. The distorted vortex beam is corrected after passing through the compensation phase screen, and the intensity distribution is also converged into a perfect doughnut. Fig. 1(b), Fig. 1(c) and Fig. 1(d) are the mode purity of the vortex beam before passing through OT, after passing through OT and after being compensated, respectively.
\begin{figure} [!htbp]
\label{FIG_1}
\centering
\includegraphics[width=0.5\columnwidth]{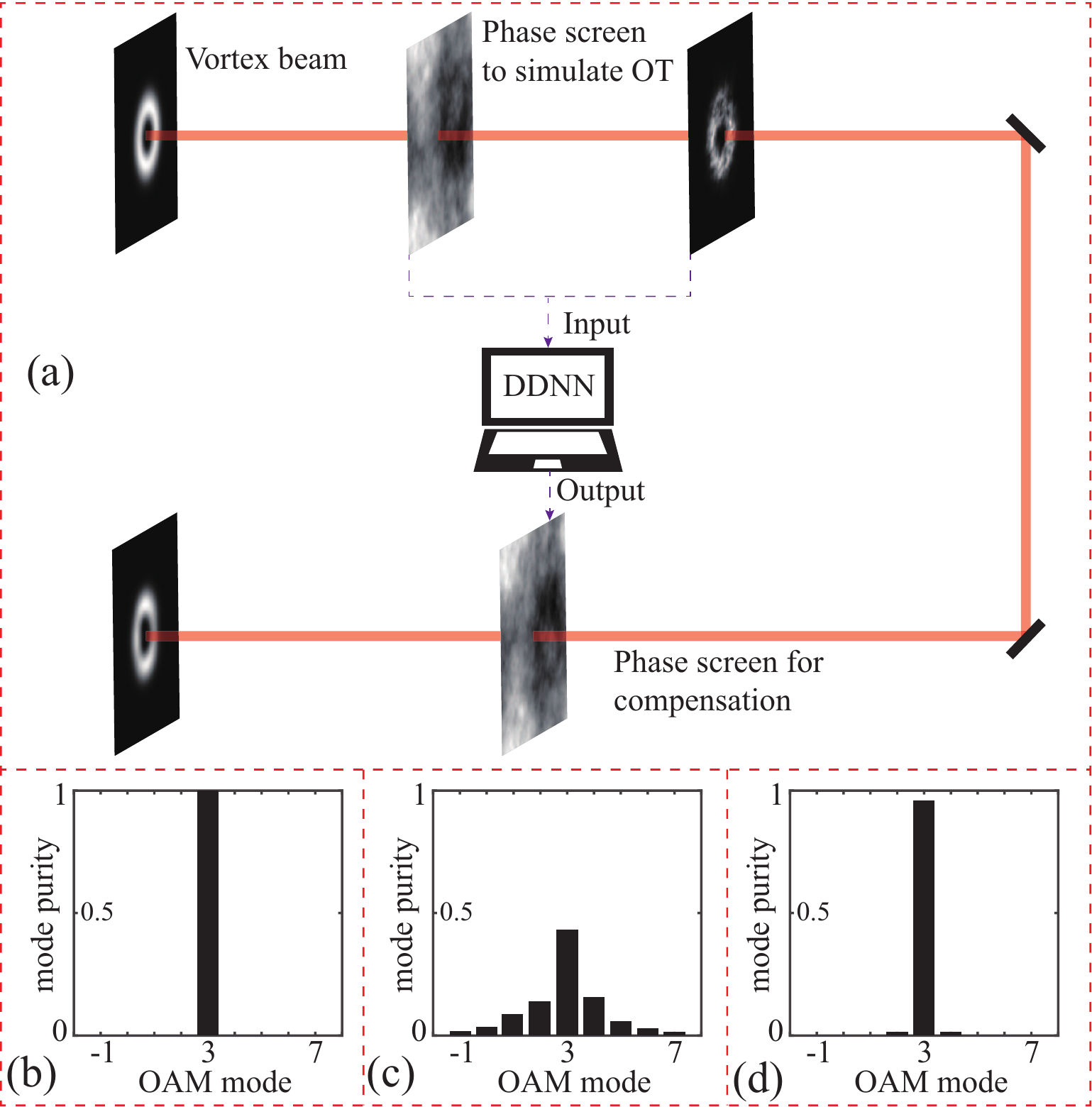}
\parbox[c]{12cm}{\footnotesize{\bf Fig. 1.} Schematic diagram of the AO compensation scheme based on DDNN. (a) The prediction and compensation of AO based on DDNN. (b) Mode purity of the original vortex beam. (c) Mode purity of the vortex beam after passing through OT. (d) Mode purity after the distorted vortex beam is compensated.}
\end{figure}

In order to simulate the effect of OT on the beam propagation, the power spectrum inversion method is used to generate separate phase thin sheet at regular intervals \cite{ISI:000488224100003}. The phase thin sheet is also called phase screen, which can adjust the phase of the beam. On the beam propagation axis, the phase disturbance caused by the fluctuation of the refractive index can be expressed as the multiplication of the input field and the phase exponential function (a phase screen). And the phase screen is independent of spatial statistics matching the index of refraction fluctuations. The spectrum of refractive index fluctuations of homogeneous and isotropic OT \cite{chang2019performance} can be expressed as:
\begin{equation} \label{eqn1}
 \begin{aligned}
   {{\Phi }_{ot}}\left( \kappa  \right)&=0.388C_{n}^{2}{{\kappa }^{-11/3}}\left[ 1+2.35{{\left( \kappa \eta  \right)}^{2/3}} \right] \\
    & \times \left[ \exp \left( -{{A}_{T}}\delta  \right)-2{{\tau }^{-1}}\exp \left( -{{A}_{TS}}\delta \right)+{{\tau }^{-2}}\exp \left( -{{A}_{S}}\delta  \right) \right],
 \end{aligned}
\end{equation}
where $C_{n}^{2}={{10}^{-8}}{{\varepsilon }^{-1/3}}{{\chi }_{T}}$ is a measure of the strength of the temperature fluctuations in the ocean. $\varepsilon $, ${{\chi }_{T}}$, and $\tau $ are the kinetic energy dissipation rate per unit mass of fluid, the dissipation rate of the mean square temperature, and the balance parameter that means the relative strength of temperature and salinity fluctuations, respectively, which ranges from ${{10}^{-10}}{{\text{m}}^{2}}/{{\text{s}}^{3}}$ to ${{10}^{-1}}{{\text{m}}^{2}}/{{\text{s}}^{3}}$, from ${{10}^{-10}}{{\text{K}}^{2}}/\text{s}$ to ${{10}^{-4}}{{\text{K}}^{2}}/\text{s}$, and from -5 to 0. $\kappa $ and $\eta $ are the spatial frequency and the Kolmogorov microscale (inner scale), respectively. ${{A}_{T}}=1.863\times {{10}^{-2}}$, ${{A}_{S}}=1.9\times {{10}^{-4}}$, ${{A}_{TS}}=9.41\times {{10}^{-3}}$, and $\delta =8.284{{\left( \kappa \eta  \right)}^{4/3}}+12.978{{\left( \kappa \eta  \right)}^{2}}$. The phase spectrum can be expressed as:
\begin{equation} \label{eqn2}
\Phi \left( \kappa  \right)=2\pi k_{0}^{2}Z{{\Phi }_{ot}}\left( \kappa  \right),
\end{equation}
where ${{k}_{0}}$ denotes the wave number of the incident beam, $Z$ is the propagation distance. So, the phase screen can be obtained by the Fourier transform:
\begin{equation} \label{eqn3}
\varphi \left( x,y \right)=F\left[ C\left( \frac{2\pi }{N\Delta x} \right)\sqrt{\Phi \left( \kappa  \right)} \right],
\end{equation}
where $F$ is the Fourier transformation, $C$ represents an $N\times N$ array of complex random numbers with 0 mean and variance 1, and $\Delta x$ is the grid spacing.

\subsection{Network structure}

\begin{figure}[!htbp]
\label{FIG_2}
\centering
\includegraphics[width=0.5\columnwidth]{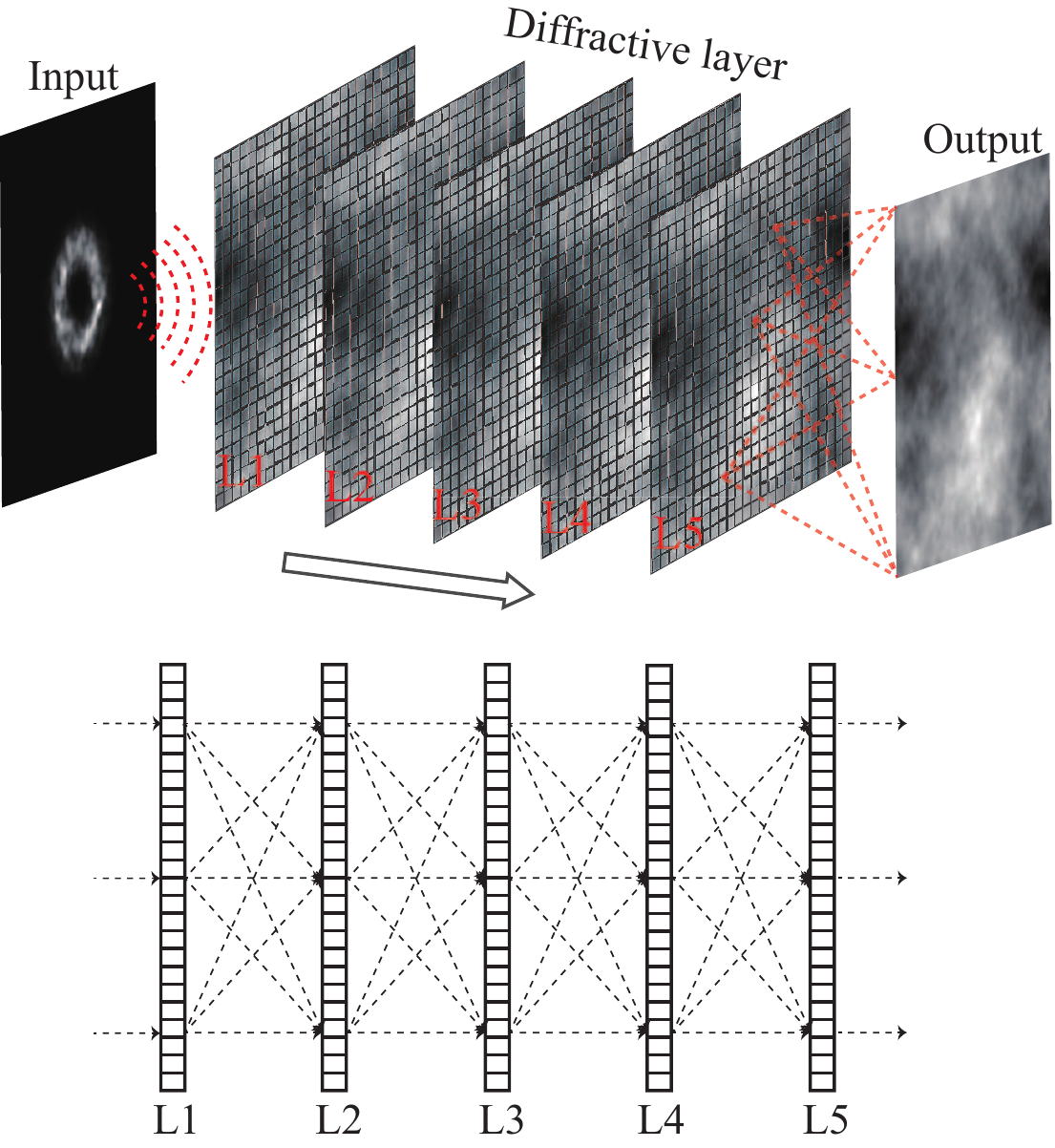}
\parbox[c]{11cm}{\footnotesize{\bf Fig. 2.} The framework of the DDNN used to compensate for wavefront distortion.}
\end{figure}

DDNN, with learning, memory and light field processing capabilities, is composed of a series of transmission layers or reflective diffraction layers. Fig. 2 shows the framework of the DDNN used to compensate for wavefront distortion. Each sensor node on the diffraction layer corresponds to a neuron in the neural network layer, and each node is fully connected to all sensor nodes in the next diffraction layer. Following the Rayleigh-Sommerfeld diffraction equation \cite{Qian:20}, a single sensor node of DDNN can be considered as a secondary source of a wave, which is composed of the following optical mode:
\begin{equation} \label{eqn4}
\omega _{i}^{N}\left( x,y,z \right)=\frac{z-{{z}_{i}}}{{{r}^{2}}}\left( \frac{1}{j\lambda }+\frac{1}{2\pi r} \right)\exp \left( \frac{j2\pi r}{\lambda } \right),
\end{equation}
where $N$ denotes the $N-th$ layer of the DDNN, $i$ represents the sensor node located at $\left( {{x}_{i}},{{y}_{i}},{{z}_{i}} \right)$ of the $N-th$ layer. $r=\sqrt{{{\left( x-{{x}_{i}} \right)}^{2}}+{{\left( y-{{y}_{i}} \right)}^{2}}+{{\left( z-{{z}_{i}} \right)}^{2}}}$, is the distance from the source to the $i-th$ sensor node in the $N-th$ layer, $j=\sqrt{-1}$, and $\lambda $ is the wavelength. The amplitude and phase of the secondary wave are determined by the input wave to the sensor nodes and its transmission coefficient $t_{i}^{N}\left( {{x}_{i}},{{y}_{i}},{{z}_{i}} \right)$. So, the output function of the $i-th$ sensor node in the $N-th$ layer can be expressed as:
\begin{equation} \label{eqn5}
s_{i}^{N}\left( {{x}_{i}},{{y}_{i}},{{z}_{i}} \right)=\omega _{i}^{N}\left( x,y,z \right)\cdot t_{i}^{N}\left( {{x}_{i}},{{y}_{i}},{{z}_{i}} \right)\cdot \sum\nolimits_{k}{s_{k}^{N-1}\left( {{x}_{i}},{{y}_{i}},{{z}_{i}} \right)},
\end{equation}
where $\sum\nolimits_{k}{s_{k}^{N-1}\left( {{x}_{i}},{{y}_{i}},{{z}_{i}} \right)}$ is the input wave superposition of all nodes of the $\left( N-1 \right)th$ layer to the $i-th$ sensor node of the $N-th$ layer. According to the optical characteristics, the transmission coefficient of the sensor node is composed of amplitude and phase, $t_{i}^{N}\left( {{x}_{i}},{{y}_{i}},{{z}_{i}} \right)=a_{i}^{N}\left( {{x}_{i}},{{y}_{i}},{{z}_{i}} \right)\exp \left( j\phi _{i}^{N}\left( {{x}_{i}},{{y}_{i}},{{z}_{i}} \right) \right)$. So, Eq. (5) can be simplified to:
\begin{equation} \label{eqn6}
s_{i}^{N}\left( {{x}_{i}},{{y}_{i}},{{z}_{i}} \right)=\omega _{i}^{N}\left( x,y,z \right)\cdot \left| A \right|\exp \left[ j\theta \left( {{x}_{i}},{{y}_{i}},{{z}_{i}} \right) \right],
\end{equation}
where $\left| A \right|$ and $\theta \left( {{x}_{i}},{{y}_{i}},{{z}_{i}} \right)$ are the relative amplitude and additional phase delay of the secondary wave caused by the input wave and transmission coefficient, respectively. According to the variable amount in the transmission coefficient, DDNN can be divided into three types: phase modulation only, amplitude modulation only and hybrid modulation. The secondary wave of the sensor node in the previous layer is diffracted to the next layer, and to the output layer finally.

The forward propagation process of vortex beam between DDNN diffractive layers can be simulated by Fresnel diffraction theory \cite{PhysRevApplied.15.014037}. The frequency domain distribution of the incident beam on the $\left( N-1 \right)th$ layer can be obtained by the Fourier transform of the light field in the spatial domain, as shown in the following equation:
\begin{equation} \label{eqn7}
U_{{{z}_{N-1}}}^{N-1}\left( {{f}_{x}},{{f}_{y}} \right)=\mathbb{F}\left( u_{{{z}_{N-1}}}^{N-1}\left( {{x}_{N-1}},{{y}_{N-1}} \right) \right),
\end{equation}
where ${{z}_{N-1}}$ is the observation plane on the propagation axis, $\mathbb{F}$ is the Fourier transform. Therefore, the output of the $\left( N-1 \right)th$ diffraction layer can be expressed as:
\begin{equation} \label{eqn8}
U_{{{z}_{N-1}}}^{N-1}\left( {{f}_{x}},{{f}_{y}} \right)=\mathbb{F}\left( u_{{{z}_{N-1}}}^{N-1}\left( {{x}_{N-1}},{{y}_{N-1}} \right)\cdot t_{{{z}_{N-1}}}^{N-1}\left( {{x}_{N-1}},{{y}_{N-1}} \right) \right).
\end{equation}
After the beam propagates to the $N-th$ diffraction layer, based on the Fresnel diffraction theory, the light field in the spatial domain can be expressed as:
\begin{equation} \label{eqn9}
u_{{{z}_{N}}}^{N}\left( {{x}_{N}},{{y}_{N}} \right)={{\mathbb{F}}^{-1}}\left( U_{{{z}_{N-1}}}^{N-1}\left( {{f}_{x}},{{f}_{y}} \right)\cdot H\left( {{f}_{x}},{{f}_{y}} \right) \right),
\end{equation}
where ${{\mathbb{F}}^{-1}}$ represents the reverse Fourier transform. $H\left( {{f}_{x}},{{f}_{y}} \right)$ is the transform function, $H\left( {{f}_{x}},{{f}_{y}} \right)=\exp \left[ jk\left( {{z}_{N}}-{{z}_{N-1}} \right) \right]\exp \left[ -j\pi \lambda \left( {{z}_{N}}-{{z}_{N-1}} \right)\left( f_{x}^{2}+f_{y}^{2} \right) \right]$, where $k=2\pi /\lambda $ is the wave number, and $\left( {{z}_{N}}-{{z}_{N-1}} \right)$ represents the separation distance between the diffractive layers.

In order to ensure the consistency of the network output results and the target output (ground truth), the error backward propagation algorithm and Adam gradient descent optimizer \cite{kingma2015adam} are used to train the DDNN. The loss function is defined as:
\begin{equation} \label{eqn10}
loss={{\left( {{x}_{o}}-{{x}_{gt}} \right)}^{2}},
\end{equation}
where ${{x}_{o}}$ and ${{x}_{gt}}$ represent the actual output of DDNN and ground truth, respectively.

\section{Performance evaluation}

In order to evaluate the compensation performance of the AO scheme based on DDNN, we utilize LabVIEW 2012 to simulate and acquire the OT phase screen and the distorted vortex beam intensity distribution images. We choose phase screens with different OT intensity levels and corresponding vortex beam intensity distributions, where $C_{n}^{2}=1\times {{10}^{-15}}{{\text{K}}^{\text{2}}}{{\text{m}}^{\text{-2/3}}}$ represents weak turbulence, $C_{n}^{2}=1\times {{10}^{-14}}{{\text{K}}^{\text{2}}}{{\text{m}}^{\text{-2/3}}}$ and $C_{n}^{2}=1\times {{10}^{-13}}{{\text{K}}^{\text{2}}}{{\text{m}}^{\text{-2/3}}}$ represent medium turbulence, and $C_{n}^{2}=1\times {{10}^{-12}}{{\text{K}}^{\text{2}}}{{\text{m}}^{\text{-2/3}}}$ represents strong turbulence. The beam propagation distance $Z$ at each turbulence intensity level is 30m. Each OT intensity level has 12,000 random phase screens, which are marked as ground truth. At the same time, the corresponding 12,000 vortex beam intensity distribution images are also recorded. For each turbulence intensity level, the training dataset includes 10,000 ground truth and 10,000 distortion intensity images, and the remaining 2,000 sets of images are used as testing dataset. The original images obtained from LabVIEW 2012 are resized to 256$\times $256, converted to gray scale images, and normalized. The training of DDNN model is implemented with the Tensorflow framework on four NVIDIA TITAN RTX GPUs. Furthermore, the DDNN model are trained for 50 epochs using the Adam optimizer with a learning rate of 0.01.
\begin{figure} [!htbp]
\label{FIG_3}
\centering
\includegraphics[width=0.6\columnwidth]{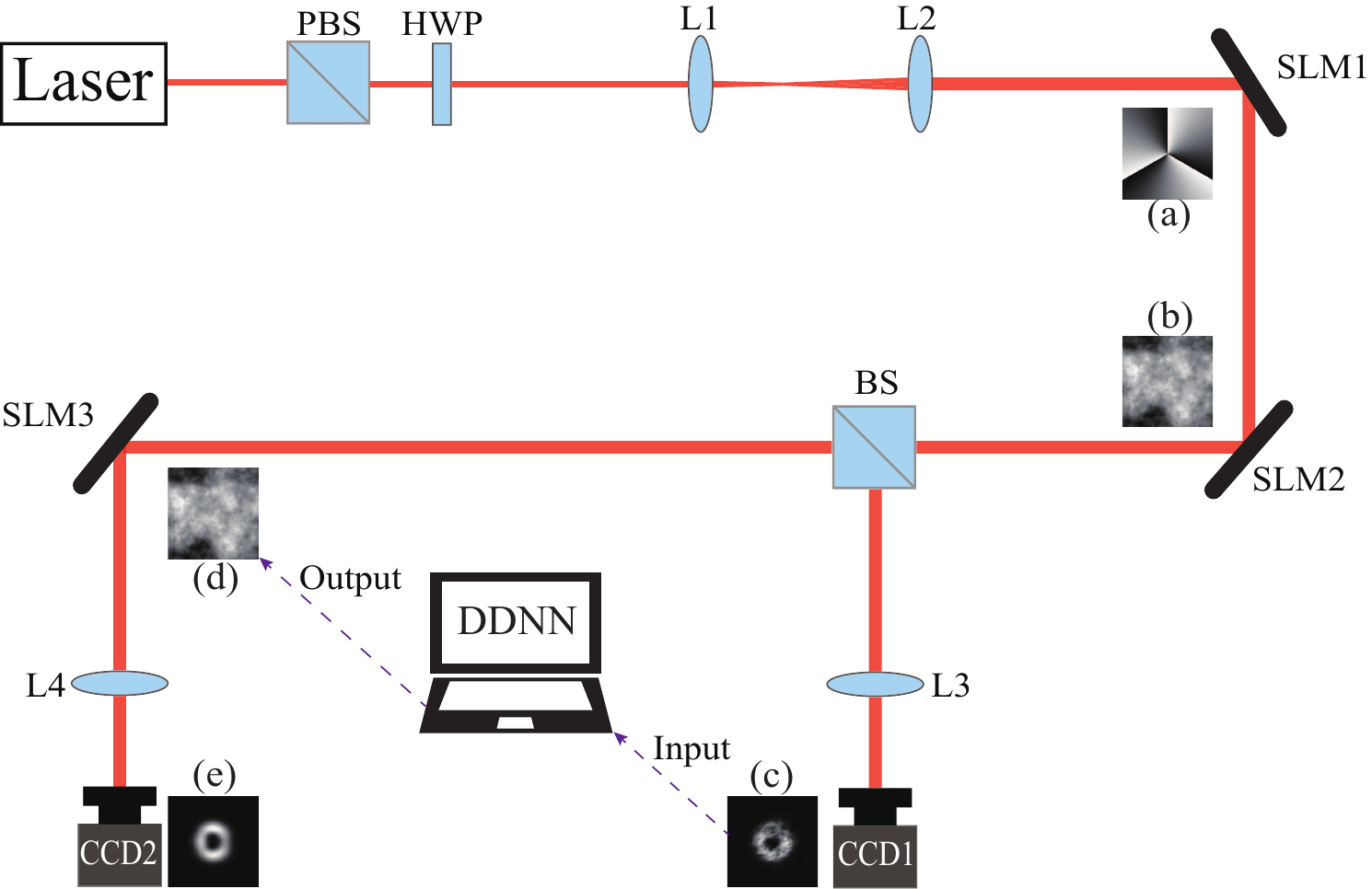}
\parbox[c]{12cm}{\footnotesize{\bf Fig. 3.} The experimental setup of AO compensation scheme based on DDNN. L1, L2: beam expander; L3, L4: Fourier lens; SLM: spatial light modulator; CCD: charge coupled device camera. (a) The OAM phase mask. (b) The phase screen simulating OT. (c) The distorted intensity pattern of vortex beam. (d) The predicted phase screen output by DDNN. (e) The intensity pattern of the vortex beam after compensation. PBS: polarization beam splitter; HWP: half-wave plate; L1, L2: beam expander; L3: Fourier lens; SLM: spatial light modulator; BS: beam splitter; CCD: charge coupled device camera.}
\end{figure}

Fig. 3 shows the experimental setup of AO compensation scheme based on DDNN. The laser (Thorlabs, HNL050R) emits a Gaussian beam with beam waist of 0.0035m and wavelength of 633nm. The Gaussian beam passes through a polarization beam splitter (PBS) (Thorlabs, CM1-PBS252) and half-wave plate (HWP) (Thorlabs, RSP1C/M) to obtain linearly polarized light beam that matches the spatial light modulators (SLM) (Hamamatsu, X10468-07). Then, the Gaussian beam is enlarged by a beam expander composed of two lenses, L1 and L2, and the focal lengths of the two lenses are 50mm and 100mm, respectively. The expanded Gaussian beam is propagated through three SLMs. L3 and L4 are Fourier lens (Thorlabs, LMR1/M, AC254-100-A), which are used to focus the light intensity distribution to CCD camera (Thorlabs, BC106N-VIS/M) in Fourier plane. The CCDs are utilized to record and save the intensity pattern of the vortex beam before and after compensation. Firstly, an OAM phase mask ($\ell =-3$ in the Fig. 3(a)) is imprinted to the SLM1. At this time, the intensity pattern of the vortex beam recorded by CCD1 is a perfect doughnut structure. Then, a phase screen, which simulates OT, as shown in Fig. 3(b), is loaded into the SLM2. After the vortex beam passes through the phase screen, its spiral wavefront is distorted. At this moment, the intensity pattern observed by CCD1 is as shown in Fig. 3(c). The distorted intensity pattern is resized to 256$\times $256, converted to gray scale images, normalized and input into the trained DDNN model. As shown in Fig. 3(d), the predicted phase screen is output by the DDNN model. Finally, the predicted phase screen is reversed phase to obtain the compensated phase screen. The compensation phase screen is uploaded to the SLM3 to compensate for the distortion caused by OT. As shown in Fig. 3(e), The intensity pattern of the compensated vortex beam is captured by the CCD2. Mode purity, also called transmission probability \cite{ZHAN2021166990}, can be used to evaluate the purity of OAM mode, which can be expressed as:
\begin{equation} \label{eqn11}
M{{P}_{m}}=\frac{{{I}_{m}}}{\sum\limits_{\ell }{{{I}_{\ell }}}},
\end{equation}
where ${{I}_{m}}$ is the light intensity of the $\ell =m$. ${{I}_{\ell }}$ denotes the light intensity of all modes at the receiver. The quality of the predicted phase screen can be quantitatively evaluated by the peak signal-to-noise ratio (PSNR) \cite{Zhao:13}.
\begin{equation} \label{eqn12}
PSNR=10{{\log }_{10}}\left( \frac{Max_{Val}^{2}}{MSE} \right),
\end{equation}
where $Max_{Val}^{2}$ represents the maximum possible pixel value of the predicted phase screen. The $MSE$ can be expressed as:
\begin{equation} \label{eqn13}
MSE=\frac{1}{N}\sum\limits_{x,y}{{{\left( {{T}^{'}}\left( x,y \right)-T\left( x,y \right) \right)}^{2}}},
\end{equation}
where $N$ is the number of pixels of the image. ${{T}^{'}}\left( x,y \right)$ and $T\left( x,y \right)$ denote the predicted phase and the ground truth, respectively.

We construct a five-layer DDNN to establish the mapping relationship between the distortion intensity pattern of vortex beam and the phase screen. With training, the difference between the predicted phase screen and ground truth is gradually eliminated. Fig. 4 shows the predicted phase and amplitude distribution of each diffraction layer after the training is completed. For the hybrid modulation type DDNN, each diffraction layer contains the predicted phase and amplitude distribution.
\begin{figure}[!htbp]
\label{FIG_4}
  \centering
  \includegraphics[width=0.7\columnwidth]{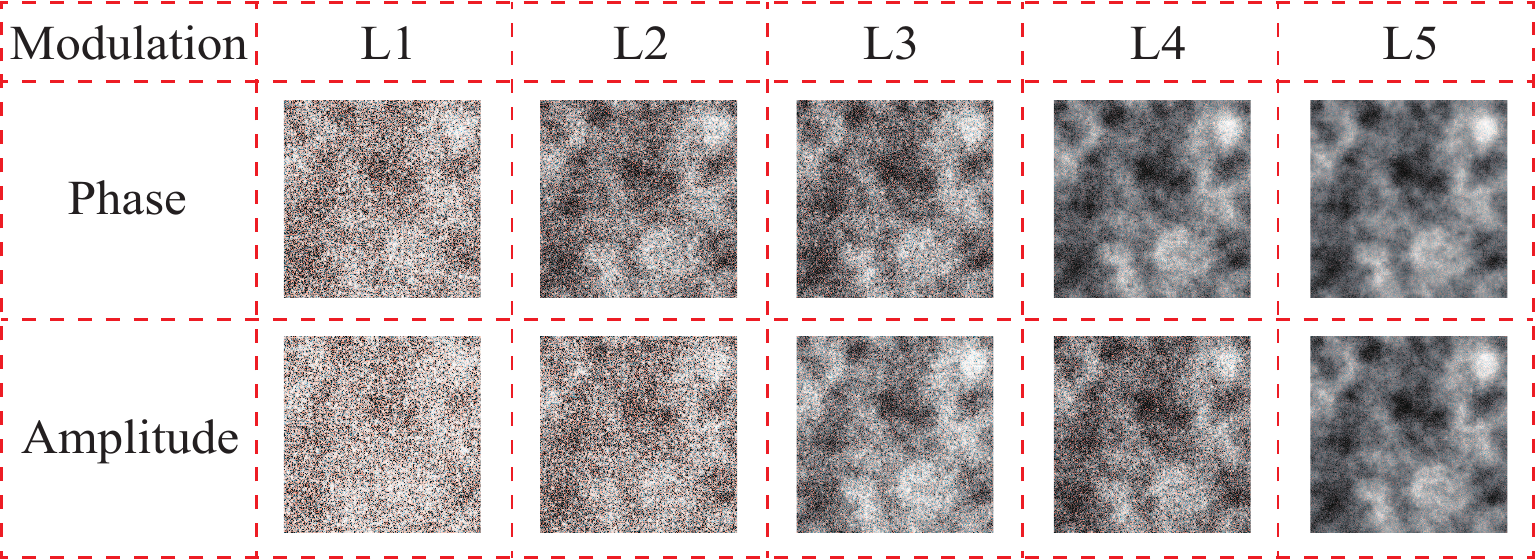}
  \parbox[c]{11cm}{\footnotesize{\bf Fig. 4.} The predicted phase and amplitude distribution of each diffraction layer.}
\end{figure}

Fig. 5 shows the ground truth at different OT intensities, the predicted phase screen output by the DDNN, and the experimental results of the intensity pattern before and after the vortex beam is compensated. When the vortex beam is not interfered by OT, its MP is 1. As the intensity of OT increases, the distortion of the vortex beam becomes more serious. The MP under strong turbulence is only 0.156, while under weak turbulence it is 0.513. The phase screen predicted by DDNN is reversed phase operation to obtain the compensated phase screen, and then loaded into SLM3, the distortion of the vortex beam is restored. The intensity pattern of the vortex beam is also restored to the doughnut structure, even under strong OT, the MP reaches 0.980. In addition, comparing the PSNR between the predicted phase screen and ground truth under different OT intensities, it can be found that the predictive ability of DDNN is not affected by the intensity of OT, which provides important support for AO compensation in strong OT environments.
\begin{figure}[!htbp]
\label{FIG_5}
  \centering
  \includegraphics[width=0.7\columnwidth]{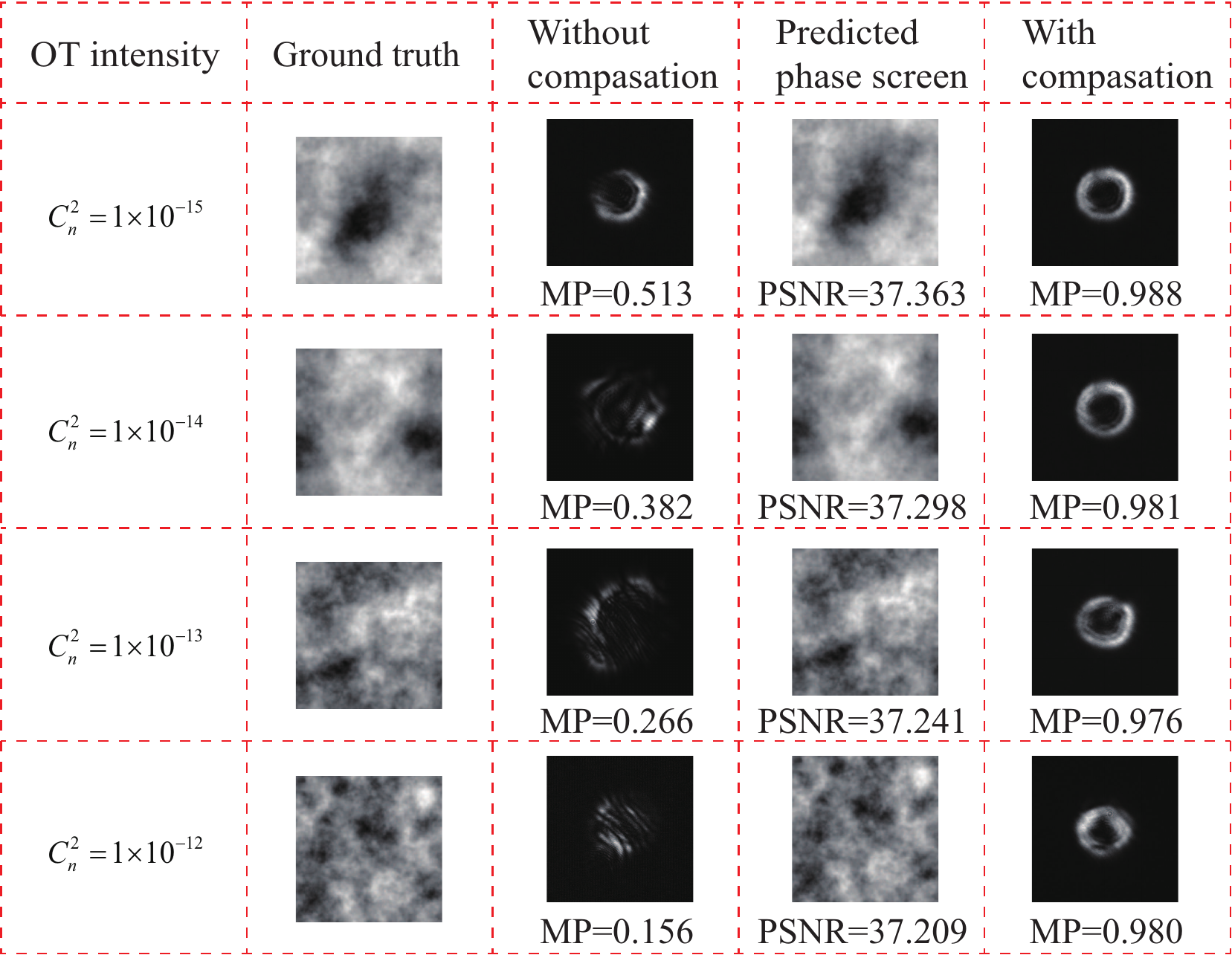}
  \parbox[c]{11cm}{\footnotesize{\bf Fig. 5.} The prediction results of DDNN under different OT intensities and the experimental results of the vortex beam intensity patterns before and after compensation.}
\end{figure}

To further evaluate the performance of the DDNN-based AO compensation scheme under strong OT, we compare the influence of different epoch numbers on the predicted phase screen output by DDNN, as shown in Fig. 6. Under strong OT, with the increase of epoch, the predicted phase screen is gradually consistent with ground truth. The PSNR reaches a maximum of 37.209 around 50 epochs, which means that the predicted phase screen is almost equivalent to the ground truth.
\begin{figure}[!htbp]
\label{FIG_6}
  \centering
  \includegraphics[width=0.7\columnwidth]{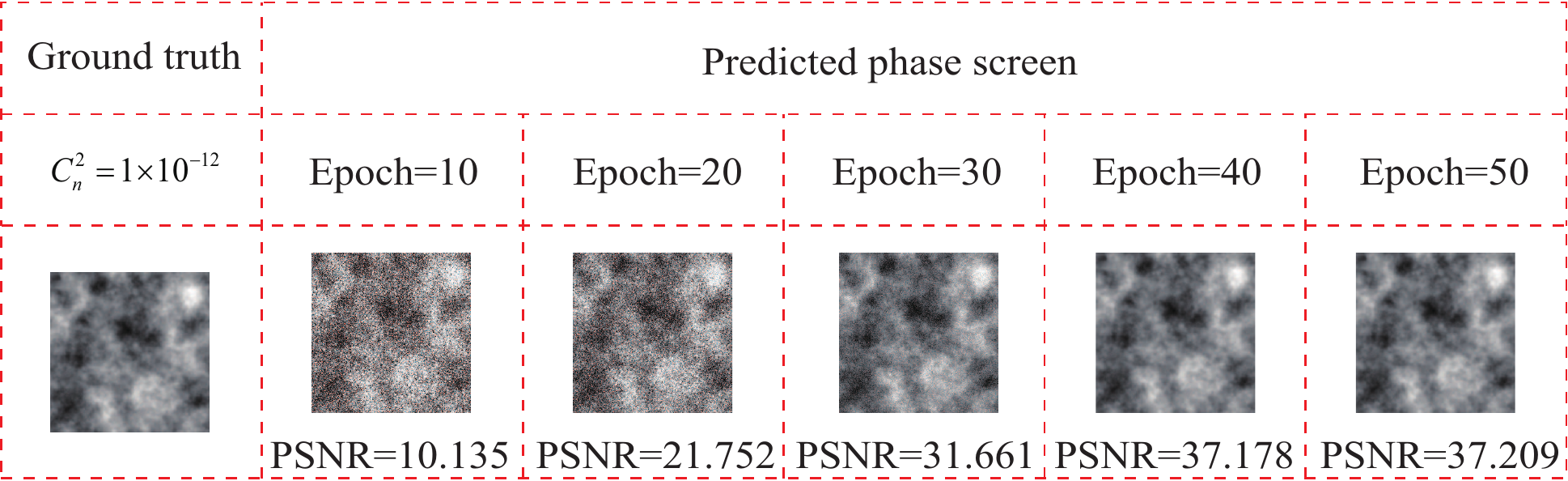}
  \parbox[c]{11cm}{\footnotesize{\bf Fig. 6.} The predicted phase screen of DDNN output under different epochs.}
\end{figure}
We use the predicted phase screen output by the DDNN under different epochs to compensate for the distortion of the vortex beam, and obtain its MP, which is recorded in Table 1. The change trend of MP with epoch corresponds to Fig 6. When the epoch is 50, the MP reaches the maximum value of 0.980, which shows that DDNN can excellently compensate the distortion of the vortex beam even under strong OT.
\begin{table}[!htbp]
	\centering
	\caption{The MP after the vortex beam is compensated under different epochs.}
	\label{tab:1}
	\begin{tabular}{cccccc}
		\hline\noalign{\smallskip}	
		Epoch & 10 & 20 & 30 & 40 & 50  \\
		\noalign{\smallskip}\hline\noalign{\smallskip}
		MP & 0.208 & 0.339 & 0.763 & 0.966 & 0.980 \\
		\noalign{\smallskip}\hline
	\end{tabular}
\end{table}

The CNN-based AO compensation schemes are basically divided into two types, one outputs the coefficients of the Zernike polynomial representing the turbulence phase, and the other outputs the phase screen directly. We upload the compensation phase screen generated by various AO methods to SLM3, record the intensity pattern of the vortex beam through CCD2, and calculate the MP. Then, we compare the MP after the vortex beam is compensated by various AO methods under different OT intensities, as shown in Fig. 7, CCN1 represents the output of Zernike coefficients, CNN2 represents the output of phase screen. Under weak OT intensity, DDNN, CNN1, CNN2 and GS algorithm achieve excellent compensation performance, while under strong OT intensity, only DDNN can well compensate the distortion of the vortex beam, MP is close to 1.
\begin{figure}[!htbp]
\label{FIG_7}
  \centering
  \includegraphics[width=0.5\columnwidth]{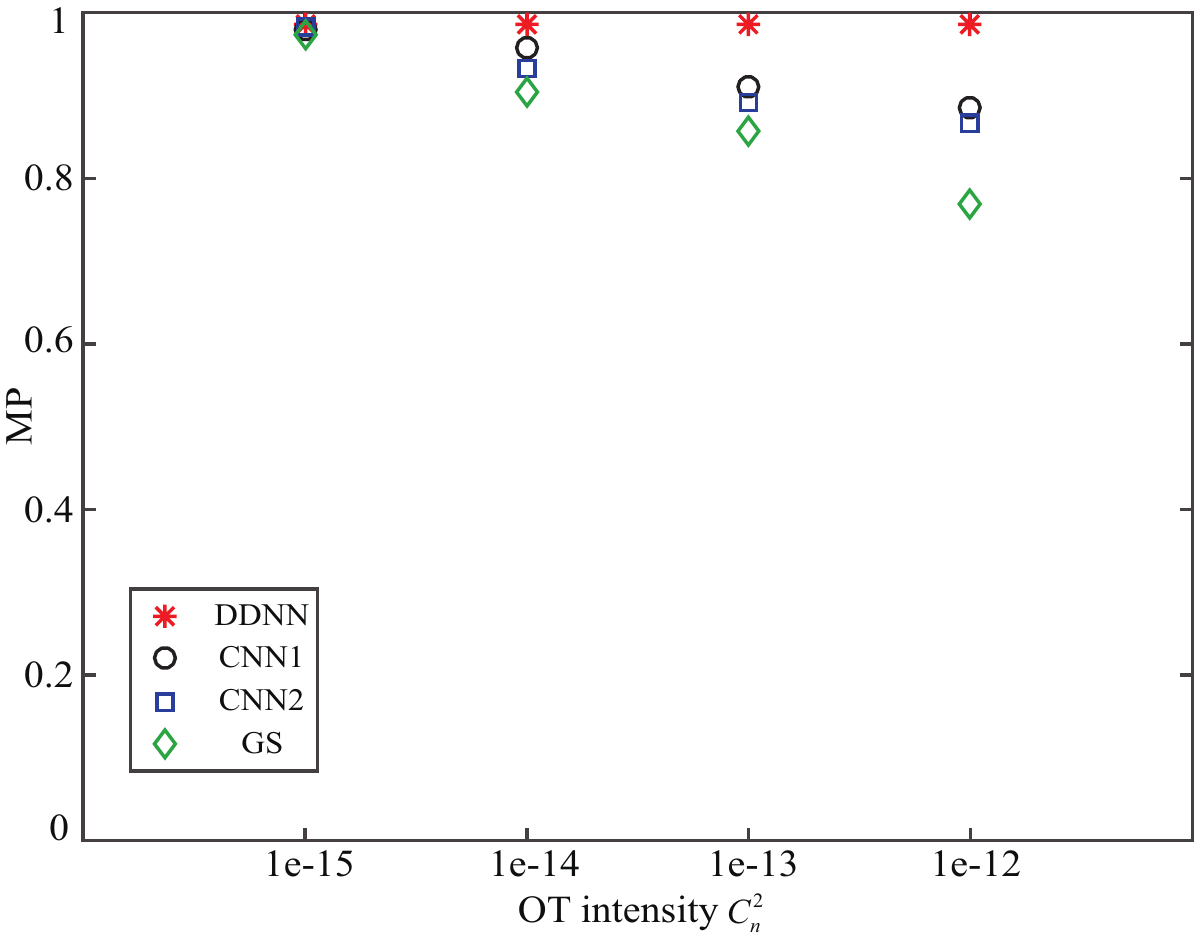}
  \parbox[c]{11cm}{\footnotesize{\bf Fig. 7.} The MP of vortex beam compensated by AO compensation scheme based on DDNN and CNN, and GS algorithm.}
\end{figure}

We record the response time of various AO methods from processing distortion information to outputting prediction results. Fig. 8 shows the response time of various AO methods under different OT intensities on the same computer. As the intensity of OT increases, the response time of the GS algorithm that needs to be iterated also increases. Under strong OT, it takes 20 seconds for the GS algorithm to complete 100 iterations, and may fall into the trouble of local optimization. We use the trained DDNN and CNN models to record the time required for prediction, so the time for datasets preparation and training is not taken into account. It can be seen from Fig. 8 that the response time of DDNN and CNN models does not change with the increase of OT intensity. The response time of the DDNN model is the least, only 0.977s. Compared with the 2.156s of CNN2, CNN1 needs to convert the predicted Zernike coefficients into the compensation phase, so its response time is 3.823s.
\begin{figure}[!htbp]
\label{FIG_8}
  \centering
  \includegraphics[width=0.5\columnwidth]{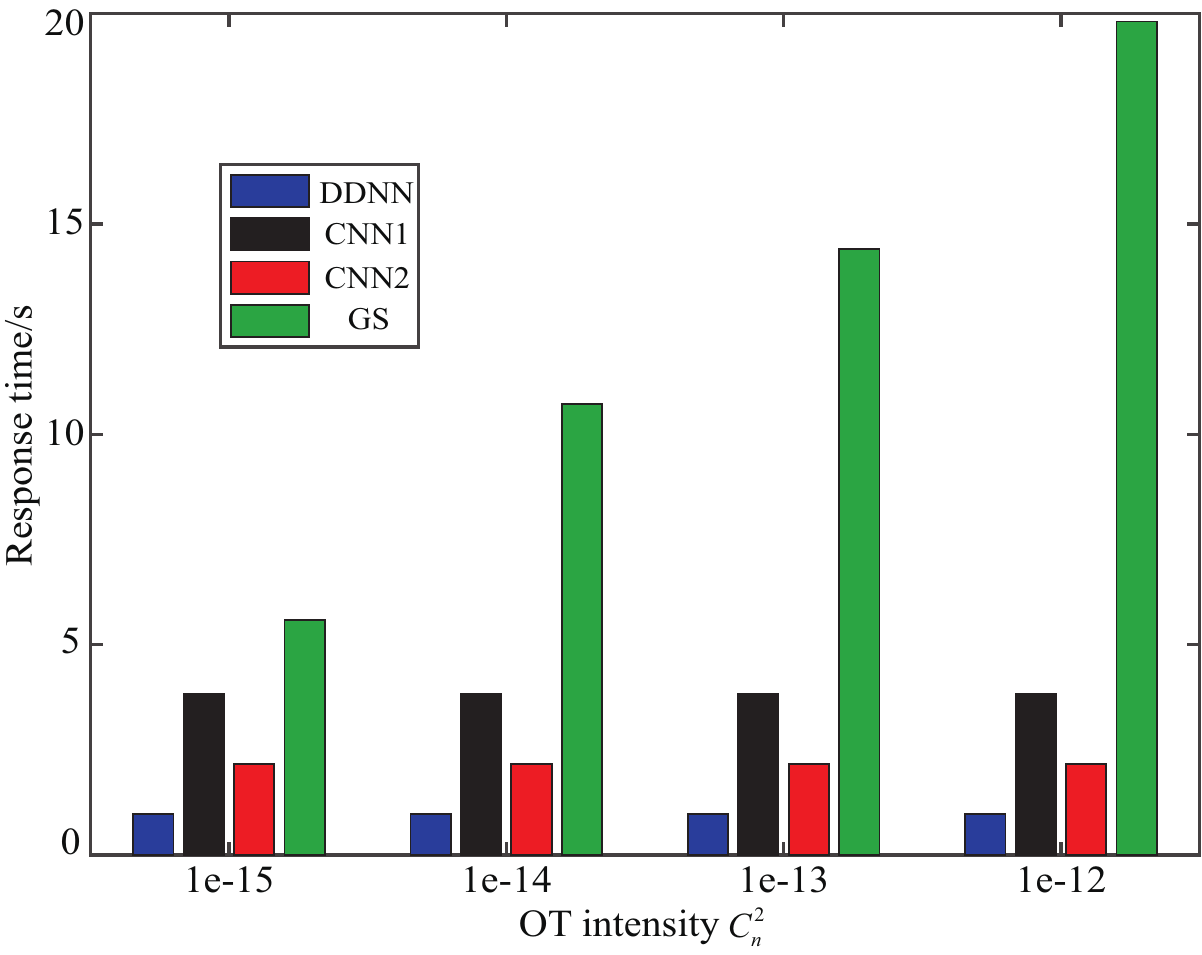}
  \parbox[c]{11cm}{\footnotesize{\bf Fig. 8.} The response time of AO compensation scheme based on DDNN and CNN, and GS algorithm.}
\end{figure}

The change of the training loss function can evaluate the closeness of the prediction results and the output results. Fig. 9 shows the change curve of the training loss of the AO scheme based on DDNN and CNN within 50 epochs. The results show that the training loss of the DDNN model has a faster convergence trend than the CNN1 and CNN2 models, and the final stable value is also lower, which means that the AO compensation scheme based on DDNN can achieve excellent performances.
\begin{figure}[!htbp]
\label{FIG_9}
  \centering
  \includegraphics[width=0.6\columnwidth]{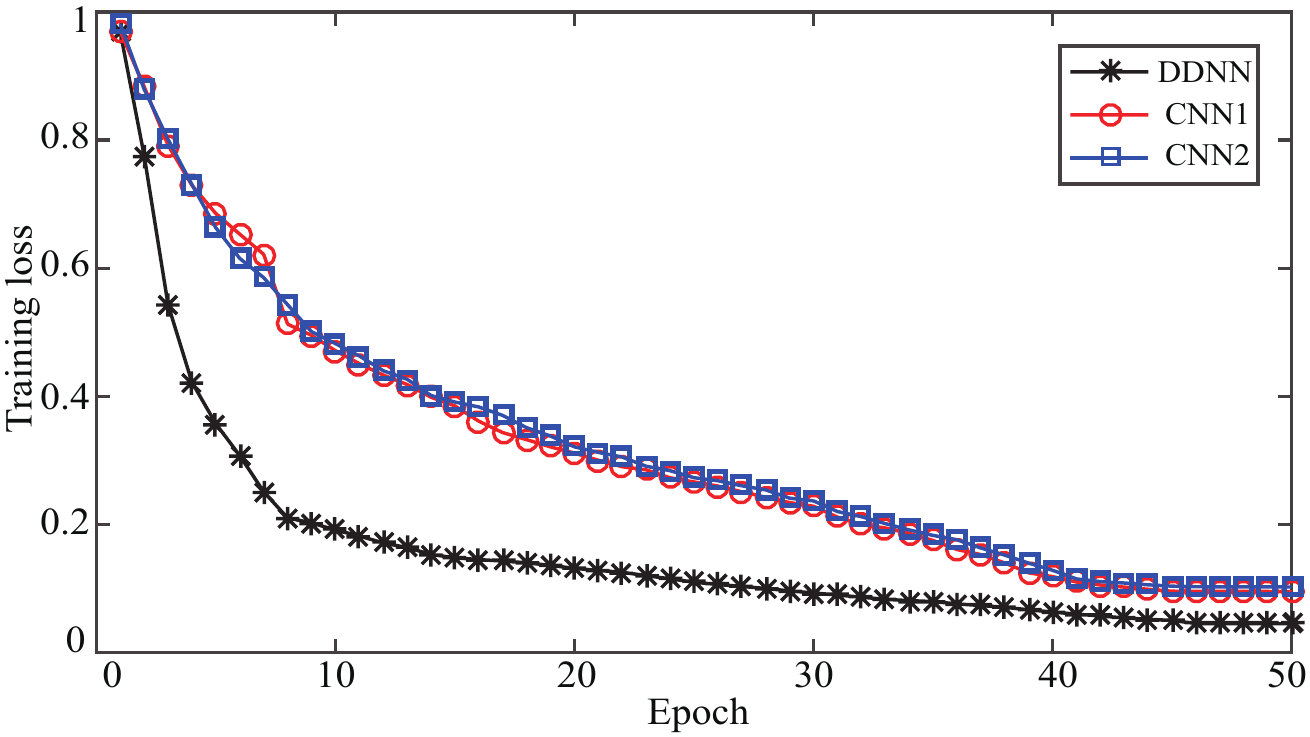}
  \parbox[c]{11cm}{\footnotesize{\bf Fig. 9.} The change curve of the training loss of the AO compensation scheme based on DDNN and CNN within 50 epochs.}
\end{figure}

\section{Conclusions}

In this work, we have proposed a DDNN-based AO compensation scheme to recover the distortion caused by OT to the vortex beam. The DDNN model has been trained to obtain the corresponding relationship between the intensity pattern of the distorted vortex beam and the phase screen. When the intensity pattern of the distorted vortex beam has been input, the trained DDNN model outputs the predicted phase screen. The predicted phase screen han been reversed phase to obtain the compensated phase screen, so as to realize the distortion compensation of the vortex beam. We have verified the ability of the DDNN-based AO scheme to compensate for the distorted vortex beam through experiments, and compared it with the CNN-based AO compensation scheme and the GS algorithm. The results show that the AO model based on DDNN can compensate the distortion caused by OT to the vortex beam, especially in a strong turbulence environment, the mode purity of the vortex beam after compensation is improved the most, reaching 0.980. In addition, the DDNN model has the advantages of low response time and fast convergence of the loss function. In short, this scheme demonstrates the combination of diffractive neural network and adaptive optics, which can compensate for the distortion of the vortex beam caused by OT in real time, and provide help for the development of the UWOC system in the future.

\section*{Funding}
National Natural Science Foundation of China (NSFC) (61871234, 62001249); Postgraduate Research and Practice Innovation Program of Jiangsu Province (KYCX200718).
\section*{Disclosures}
The authors declare that there are no conflicts of interest related to this paper.

\end{document}